%%%%%%%%%%%%%%%%%%%%%%%%%%%%%%%%%%%%%%%%%%%%%%%%%%%%%%%%%%%%%%%%%%%%%%
% APPARENT WAVE FUNCTION COLLAPSE CAUSED BY SCATTERING
% Published in Found. Phys. Lett., 6, 571-590 (1993)
% Submitted April 19, 1993.
% By Max Tegmark
% Email: max@mppmu.mpg.de
% http://www.mpa-garching.mpg.de/~max/collapse.html (faster from Europe)
% http://astro.berkeley.edu/~max/collapse.html      (faster from the US)
%%%%%%%%%%%%%%%%%%%%%%%%%%%%%%%%%%%%%%%%%%%%%%%%%%%%%%%%%%%%%%%%%%%%%%
% This is a plain TeX file (not LATeX)
% ;-)
%%%%%%%%%%%%%%%%%%%%%%%%%%%%%%%%%%%%%%%%%%%%%%%%%%%%%%%%%%%%%%%%%%%%%%

\magnification = 1100
\nopagenumbers

\def\inc#1{\hbox{\global \advance#1 1}}
\countdef\eqnr=1    \countdef\refnr=2
\eqnr=1             \refnr=1

\countdef\GRW=3
\countdef\JoosZeh=4
\countdef\Everett=5
\countdef\Ellis=6
\countdef\Wigner=7
\countdef\Kim=8
\countdef\Squires=9
\countdef\Rae=10
\countdef\Hawking=11
\countdef\Jackson=12

\countdef\GRWeq=21
\countdef\SmatrixEq=22
\countdef\RhoHitEq=23
\countdef\MasterEq=24
\countdef\WignerHitEq=25
\countdef\FokkerPlanckEq=26
\countdef\PhatrEq=27
\countdef\gOneEq=28
\countdef\gTwoEq=29
\countdef\sEq=30

\def\nexteq{\inc{\eqnr}}    
\def\nextref{\inc{\refnr}} 
\def\enr{\number\eqnr\nexteq}   
\def\rnr{$\number\refnr$\nextref} 
\def\eq{\eqno(\enr)}            
\def\ref{[\rnr]}
\def\refn{\item{\rnr.}}

\def\vx{{\bf x}} \def\vy{{\bf y}} \def\vz{{\bf z}}  
\def\vu{{\bf u}} \def\vr{{\bf r}}
\def\vp{{\bf p}} \def\vk{{\bf k}} \def\vq{{\bf q}}
\def\vo{{\bf x_1}} \def\vt{{\bf x_2}} \def\vQ{{\bf Q}} 
\def\vxy{(\vx,\vy)}  \def\vyx{(\vy-\vx)}
\def\ak{a_{{\bf k}}} \def\pk{P_{{\bf k}}} 
\def\l0{\lambda_0}    \def\leff{\lambda_{eff}}  
   \def\akh{\hat a_{\vk}}
\def\aksh{\widehat{a^*_{\vk}}} \def\ri{\rho_i\vxy} 
\def\rf{\rho_f\vxy} \def\rh{\rho\vxy} 
\def\psii{{\psi}_i}  \def\psif{{\psi}_f}
\def\tpc{(2\pi)^3} \def\ootpc{{1 \over (2\pi)^3}} 
\def\oh{\hbox{{$1 \over 2$}}} \def\ihbar{\hbox{{$i \over\hbar$}}}
 \def\pkh{\hat P_{\vk}} \def\pkhyx{\pkh\vyx}
\def\ph{\hat P} \def\phyx{\ph\vyx}
  
\def\vnrm{V^{-1/2}} 
\def\sq{\sum_{\vq}}

\def\sxx{\biggl({\sin x \over x}\biggr)}

\def\dsdo{d\sigma/d\Omega}

\def\ftp{f(\theta,\varphi)}
\def\ct{\cos\theta}   \def\st{\sin\theta}
\def\cx{\cos x}   

\def\sxx{{\sin x \over x}}

\def\ddpi{{\partial \over {\partial p_i}}}
\def\ddpj{{\partial \over {\partial p_j}}}
\def\ddxi{{\partial \over {\partial x_i}}}
\def\ex2{\left< x^2 \right>}
\def\ybra{\left<\vy\right|}  \def\xket{\left|\vx\right>}
\def\pkbra{\left<\vp'\vk'\right|} \def\pkket{\left|\vp\vk\right>}
\def\z3{\zeta(3)}
\def\toz3{\hbox{{$2 \over \z3$}}}
\def\izi{\int_0^{\infty}}

\def\soi{\sum_{n=1}^{\infty}}
\def\streck{\noalign{\vskip2pt\hrule\vskip2pt}}
\def\Dx{\Delta x} \def\Dp{\Delta p} \def\Dv{\Delta v}
\def\tr{\hbox{tr}\>}
\def\cm{{\rm cm}} \def\mm{{\rm mm}}  \def\nm{{\rm nm}}  \def\km{{\rm km}}
\def\A{{\rm A}}   \def\m{{\rm m}}
\def\s{{\rm s}}   \def\yrs{{\rm yrs}}
\def\ie{{\it i.e.}}
\def\qlist{\noindent\parshape 2 0.5truecm 12.1truecm 1.2truecm 11.4truecm}
\def\Td{T^{\dagger}}
\def\crr{\cr\noalign{\vskip 4pt}}
\def\equiv{:=}
\def\beginsection#1{\bl\bl\goodbreak{\noindent\bf#1}\bl}

%%%%%%%%%%%%%%%%%%%%%%%%%%%%%%%%%%%%%%%%%%%%%%%%%%%%%%%%%%%%%%%%%%%%

\hsize=12.6truecm
\vsize=19.88truecm
\hoffset=2.00truecm
%\voffset=0.78truecm
%\voffset=3.53truecm
\voffset=1.838truecm
\baselineskip=0.423truecm
\def\bl{\vskip0.423truecm}
\def\ind{\noindent\hskip1.8truecm}
% \raggedbottom
%\parskip=0.2truecm

\noindent
{\it Foundations of Physics Letters, Vol. 6, No. 6, p. 571-590 (1993)}
\vskip0.799truecm
%{\hfill}
%\vskip1.222truecm

\noindent
{\bf APPARENT WAVE FUNCTION COLLAPSE}

\noindent
{\bf CAUSED BY SCATTERING}
  
\bl\bl\bl
\ind
Max Tegmark

\bl\ind 
{\it Department of Physics, University of California}

\ind
{\it Berkeley, California 94720}

\bl\bl\bl\ind
Received April 19, 1993; revised June 10, 1993

\bl\bl

\noindent
Some experimental implications of the recent progress
on wave function collapse are calculated. Exact results are derived for the
center-of-mass wave function collapse caused by random scatterings and
applied to a range of specific examples.  The results show
that recently proposed experiments to measure the GRW
effect are likely to fail, since the effect of naturally
occurring scatterings is of the same form as the GRW
effect but generally much stronger. The same goes for attempts to
measure the collapse caused by quantum gravity as
suggested by Hawking and others.  The results also
indicate that macroscopic systems tend to be found in
states with $\Delta x\Delta p = \hbar/\sqrt 2$, but
microscopic systems in highly tiltedly squeezed states with $\Delta
x\Delta p \gg \hbar$.

\bl

\noindent
Key words: decoherence, collapse, measurement, scattering.

\beginsection{1. INTRODUCTION}

The problem of how to interpret measurement in quantum mechanics has
caused intense debate ever since 1926 and shows little sign
of abating. A whole slew of interpretations have been proposed
and can be divided into two main categories: collapse theories
and non-collapse theories. 

In the former category, one of the most
successful to date is that proposed by Ghirardi, Rimini and Weber (GRW)
in 1986 \GRW=\refnr\ref, which shows that both micro- and
macroscopic systems can be described by the same dynamical equation
providing that an extra term is added to the Heisenberg equation of
motion for the density matrix:
\GRWeq=\eqnr
$$\dot\rho = -\ihbar[H,\rho] - \Lambda(\rho-T[\rho])\eq$$
They show that if this ad hoc third term is added and if
$T[\rho]$ is chosen in a particular way that singles out the
position representation as special, then the usual
problems regarding superpositions of macroscopic systems disappear
if $\Lambda$ and a second parameter are chosen appropriately. This theory
has subsequently been generalized.

In the second category of theories, perhaps the most radical is the
one proposed by Everett,  Wheeler, Cooper, DeWitt, 
and others [\rnr\Everett=\refnr\advance\refnr 4-\rnr] between
1957 and 1970, which shows that even if one assumes that the wave function
containing the observer evolves causally according to the
Schr\"odinger equation, the observer will subjectively  {\it
experience} wave function collapse. 
Zeh, K\"ubler, Joos, Machida, Namiki, Zurek, Unruh, Cini, Peres, Partovi, 
Gallis, Fleming, Hartle and others [\rnr\advance\refnr
2\JoosZeh=\refnr\advance\refnr 9-\rnr] have strengthened this
position by showing  that for macroscopic objects, their inevitable
interaction with the environment leads to a dynamic reduction of
the density matrix (what is widely known as wave function collapse)
and superselection rules. These superselection rules tend to favor
``classical" states, and explain why we never experience say spatial
superpositions of cars or superpositions of living and dead cats.
Hence this interaction with the environment shows why the position
operator and its eigenstates play such an important role
in our perception of the world, even though the position operator is a priori
merely one out of a family of infinitely many self-adjoint operators.
This will be referred to as the {\it decoherence effect}.

Unfortunately, rather scant attention has yet been given to the
experimental implications of the decoherence effect and to actual physical
parameters. This paper addresses such practical issues, focusing on
scattering. Recently, experiments have been proposed [\Squires=\refnr\rnr,
\Rae=\refnr\rnr] to try to detect the GRW effect, but this paper shows that
such experiments are likely to fail, since a GRW effect with the parameters
originally proposed would be entirely drowned out by environmental noise. 

In Sec. 2, the effect of a single scattering is calculated and shown
to damp the off-diagonal elements in the reduced spatial density
matrix by a factor that is simply the Fourier transform of the
probability distribution for different momentum transfers. In Sec.
3, the Heisenberg and Wigner equations of motion are modified to
incorporate these usually neglected effects.  In Sec. 4 the
results are applied to a variety of cases of physical interest and
compared to the predictions of the GRW theory and quantum gravity. It
is seen that scattering and the GRW effect have almost identical
effects on the reduced density matrix, although the interpretations
are completely different. Finally, Sec. 5 contains a brief
discussion of what interpretational problems of quantum mechanics the
decoherence approach does and does not solve.

\beginsection {2. THE EFFECT OF A SINGLE SCATTERING}

As an introduction to the calculations in this section, consider
the following simple example of decoherence: A spin-${1\over 2}$ 
silver atom
is prepared with its spin in the x-direction, and then somebody measures its
spin component in the z direction without telling us the result.
This changes our density matrix for the atomic spin from
$\rho_i$ to $\rho_f$, where in the z-representation
$$\rho_i = \left(\matrix{1/2&1/2\cr1/2&1/2}\right)
  \quad\hbox{ and }\quad
  \rho_f = \left(\matrix{1/2&0\cr0&1/2}\right).$$
Thus the density matrix is reduced from describing a pure state to
a mixed state. This prediction is common to all interpretations of
quantum mechanics, but arrived at in two conceptually very different ways:
Collapse-theories postulate that the time-evolution of the
wave function of the universe is not governed by the Schr\"odinger equation
during measurement, but changes discontinuously and non-causally so that
afterwards the spin really is up or down in the z-direction - we just
do not know which. 
Non-collapse theories compute the same density matrix $\rho_f$ by
letting the total system of observer and observed evolve according
to the Schr\"odinger equation with a Hamiltonian such that they
become perfectly correlated, and take a partial trace over the
observer degrees of freedom to obtain $\rho_f$. These two
incompatible viewpoints are often referred to as Heisenberg reduction and Von
Neumann reduction, respectively. (A detailed discussion of these matters is
given by Everett {[\number\Everett]} and Kraus \ref.) Here we will
adopt the latter approach, and refer to it as the {\it decoherence approach}. 

Although it is often convenient to
treat particles as isolated systems, we all know that this is merely an
approximation. 
Occasionally a photon from the
sun scatters off of our ``isolated system". More difficult to shield
experiments from are muons created by cosmic rays, cosmic
neutrinos, the 300K blackbody radiation from our surrounding and
radiation from traces of radioactive isotopes
in the materials that our measurement apparatus is made of. All these
events change the density matrix of our particle. Joos and Zeh have studied
such scattering effects in the macroscopic limit by ignoring recoil
$[\number\JoosZeh]$, whereas the following treatment applies also to
microscopic systems.

Let us chose as our system a nonrelativistic particle of
mass $m$ whose location is described by a density matrix
$\rho$ and whose time-evolution would be governed by a Hamiltonian $H$ if it
were truly isolated from its environment. Let the inevitable interaction
with the environment be given by an interaction Hamiltonian $H_{INT}$.
In this paper we will limit our attention to the special class of interactions
with the environment that can be treated as isolated scattering processes.
By this we mean that $H_{INT} \neq 0$ only during time
intervals much shorter than the dynamical timescale of the system we are
studying, so that we can approximate the change in $\rho$ as instantaneous
and given by a transition matrix $T$,
$$\rho^T_i \to \rho^T_f \equiv T\rho^T_i \Td,$$
where $\rho^T$ is the density matrix for the total system of our particle
and an external particle that scatters off of it. 
This is normally a good
approximation when our system interacts with a rapidly moving 
particle in its vicinity. For instance, it takes a photon only about
$10^{-18}$ seconds to traverse an atom. 

We will make the following
assumptions about the T-matrix and the initial data:

\bl
{\bf Assumption (I)} $T$ conserves energy and momentum. (This is
equivalent to $T$ being invariant under temporal and spatial
translations.)
Let $\pkket$ denote the state where our system has momentum $\vp$ and
the incident particle has momentum $\vk$. Then (I) implies that 
\SmatrixEq=\eqnr
$$\pkbra T\pkket =
\delta(\vp'+\vk'-\vp-\vk)a_{\vp\vk}(\vp'-\vp),\eq$$ 
where $a_{\vp\vk}(\vq)$ is the probability amplitude for the momentum
transfer to our system to be $\vq$. This function is independent of
time by energy
conservation but may depend on both $\vp$ and $\vk$.

\bl
{\bf Assumption (II)} The function $a_{\vp\vk}$ is independent of
$\vp$, i.e
of the motion of our system. Hence we will write it as $\ak$.
(This is a good approximation if the velocity of the incident particle
is much greater than the velocity spread in $\rho$).  

\bl
{\bf Assumption (III)} The incident particle is in a momentum
eigenstate or an incoherent mixture of momentum eigenstates.
(The linewidth must be much
smaller than the wavelength. For the photons we observe, the
linewidth is typically less than 1\% of the wavelength.)
\bl

To avoid normalization
problems, let us first restrict ourselves to $L^2$ functions with
periodic boundary conditions on a cube of volume V. Unless otherwise
specified, all integrals below are to be taken over this cube and all
sums are to be taken over the discrete set of vectors
$$\Omega\equiv\lbrace(2\pi/V^{1/3})(n_x,n_y,n_z)| n_x, n_y, n_z
\hbox{ integers}\rbrace.$$
To conserve probability, $\ak$ must be
normalized so that $\pk(\vq)\equiv{|\ak(\vq)|}^2$ is a
probability distribution over $\vq$, \ie
$$\sq  \pk(\vq) = 1.$$
(In the literature, $T$ is often normalized so
that this sum equals the total cross section $\sigma$ instead.
We will take account of the cross section in Sec. 3.)

Equation (\number\SmatrixEq) shows that in the position representation, $T$
transforms the state $\pkket$ from $e^{i\vp\cdot\vo}e^{i\vk\cdot\vt}$ into
$$\sq\ak(\vq)e^{i(\vp+\vq)\cdot\vo}e^{i(\vk-\vq)\cdot\vt} =
  e^{i\vp\cdot\vo} e^{i\vk\cdot\vt} \akh(\vt-\vo),$$
where $\akh$ is the discrete Fourier transform of $\ak$, 
so for a normalized initial two-particle wave function $\psii(\vo,\vt)
= \phi(\vo)\vnrm e^{i\vk\cdot\vt}$,  we have by linearity that
$$
\eqalign{
T\psii(\vo,\vt)
&= T\bigl(\phi(\vo)\vnrm e^{i\vk\cdot\vt}\bigr) 
= \phi(\vo) \vnrm e^{i\vk\cdot\vt}\akh(\vt-\vo)\crr
&= \psii(\vo,\vt) \akh(\vt-\vo)
}$$
for any one-particle wave function $\phi(\vo)$.
Let us use the notation $\rh \equiv\ybra\rho\xket$ for density
matrices and units where $\hbar = 1$. 

\bl
{\bf Theorem} 
$\rf = \ri \pkhyx$, where $\pkh$ is the
Fourier transform of $\pk\equiv{|\ak|}^2.$

\bl
{\bf Proof}
The reduced density matrix $\rho$ for our particle is obtained
by taking a partial trace of the density matrix $\rho^T$ of the two-particle
system, so
$$\ri = \int \psii^*(\vx,\vz) \psii(\vy,\vz) d^3z =
        \int\phi^*(\vx)\phi(\vy){1 \over V} d^3z = 
\phi^*(\vx)\phi(\vy),$$ 
and by the above,
$$\eqalign{
\rf 
&= \int \psif^*(\vx,\vz) \psif(\vy,\vz) d^3z\crr
&= \int {\bigl(\psii(\vx,\vz)\akh(\vz-\vx)\bigr)}^* 
     {\bigl(\psii(\vy,\vz)\akh(\vz-\vy)\bigr)} d^3z\crr
&= \int\phi^*(\vx)\phi(\vy){1 \over V} \akh^*(\vz-\vx) \akh(\vz-\vy) d^3z\crr
&= \ri {1 \over V}\int \akh^*(\vz-\vx) \akh(\vz-\vy) d^3z.
}$$
The not very elegant $V$ made its last appearance in this paper on the
previous line. Letting it approach infinity and working with Fourier
transforms instead of Fourier series from here on, the 
last equation turns into 
$$\rf = \ri \ootpc\int \akh^*(\vz-\vx) \akh(\vz-\vy) d^3z.$$
By substituting $\vu =\vz-\vy$ and using ${\akh}^*(\vu)=\aksh(-\vu)$, 
we obtain
$$\eqalign{
\rf 
&= \ri \ootpc\int \aksh(\vy-\vx-\vu) \akh(\vu) d^3u\crr
&= \ri \ootpc\bigl(\aksh\star\akh\bigr) (\vy-\vx).
}$$
Using the convolution theorem 
$\hat f \star \hat g = \tpc\widehat{fg}$ 
now yields
$$\rf = \ri \ootpc\tpc\widehat{\ak^*\ak}(\vy-\vx) = \ri \pkhyx,$$
which completes the proof for the case where our system is
initially in a pure state, {\ie}, where $\ri$ can be written in the
form $\phi^*(\vx)\phi(\vy)$. Since an arbitrary density matrix can
be written as a sum of density matrices for pure states, the proof
for the general case follows directly from superposition.

\bl
{\bf Corollary} If the incident photon is described by a
density matrix diagonal in the momentum representation,
say an incoherent superposition of plane waves with 
the momentum probability distribution given by
$\mu(\vk)$, then
\RhoHitEq=\eqnr
$$\rf = \ri \phyx,\quad\hbox{where } P(\vq)
\equiv \int\pk(\vq)\mu(\vk)d^3k.\eq$$
This again follows directly from superposition.

Thus we see that the net result of this interaction of our
``isolated system" with the outside world is simply to
multiply its spatial density matrix by a function. 
Note that this function does not depend on
the complex amplitude $\ak$ itself, but only on its squared
modulus, the probability distribution for different
momentum transfers. The latter is uniquely determined by
the differential scattering cross section
$\sigma(\theta,\varphi)$, so $\sigma$ is the only
physical input we will need to calculate how $\rho$
evolves over time.

Before we turn to calculating $\ph$ for specific physical
examples, let us make a few observations about Eq. (\number\RhoHitEq) 
that are valid for an arbitrary probability distribution $P$. It is
straightforward to prove the following:

{\bf Observation (I)}
$|\ph(\vx)| \leq 1$, with equality if $\vx=0$ or 

\noindent 
$P(\vq)=\delta(\vq-\vq_0)$.

{\bf Observation (II)}
$\tr\rho_f = \tr\rho_i$

{\bf Observation (III)}
$\tr\rho_f^2 \leq tr\rho_i^2$, with equality iff 
$\rho_i$ is diagonal or $P(\vq) = \delta(\vq-\vq_0)$.

{\bf Observation (IV)}
$\ph(\vx) \to 0$ as $|\vx| \to \infty$ if $P$ is an integrable function.

{\bf Observation (V)}
If we define $\vQ$ to be the mean and $S$ to be the covariance matrix
of the probability distribution $P$, then 
$$\ph(\vx) = 1 - iQ_mx_m - \oh(Q_m Q_n+S_{mn})x_mx_n + O(|\vx|^3).$$
(Repeated indices are to be summed over, from $1$ to $3$.)

Observation (I) tells us that all off-diagonal elements of
the density matrix will be damped if there is any
uncertainty in the outcome of the scattering.
Observation (II) simply shows that probability is conserved,
whereas (III) says that the density matrix generally becomes
less pure - we recall that $tr \rho^2=1$ for pure
states, whereas $tr \rho^2$  takes its (nonnegative)
minimum value for states of which we have zero knowledge.
(IV), which is known as Riemann-Lebesgue's Lemma,
tells us that for most physically realistic cases, the
density matrix elements very far from the diagonal get almost
entirely damped out. (V) gives us a good grip on how the density
matrix changes near the diagonal, which will prove useful in the
following section.

\beginsection{3. THE MODIFIED HEISENBERG AND WIGNER EQUATIONS OF MOTION}

Above we derived the effects of a single scattering event. Now we
will show that exposure to a constant particle flux 
amounts to a simple modification of the Heisenberg equation of motion for
our density matrix $\rho$.

Let $\sigma$ denote the total scattering cross section and $\Phi$ the
average particle flux per unit area per unit time. 
We know from experiment that the temporal distribution of scattering events 
is well modeled by a Poisson process with intensity 
$\Lambda \equiv \sigma \Phi$. Since the probability of one scattering occurring
during the infinitesimal time interval $dt$ is $\Lambda dt$ and that for no
scattering is $1-\Lambda dt$, by Eq. (\number\RhoHitEq) we would get 
$$\rho(\vx,\vy,t+dt) = 
\rho(\vx,\vy,t)\phyx\Lambda dt + \rho(\vx,\vy,t)(1-\Lambda dt),$$
{\ie} $\dot\rho(\vx,\vy,t) = -\Lambda(1-\phyx)\rho(\vx,\vy,t)$
if our scattering were the only process that changed $\rho$. 
Since $\rho$ is also changed by its normal Sshr\"odinger time evolution, 
we obtain the following master equation:
\MasterEq=\eqnr
$$\dot\rho = -\ihbar[H,\rho] - \Lambda(\rho-T[\rho]), \eq$$
where
$$\ybra T[\rho]\xket \equiv \phyx\rh.$$  
We notice that this is of the
same form as the GRW equation (1), and soon we will indeed see that the
$T[\rho]$ that we have derived is quite similar  that postulated by GRW. They
postulate that  $\ybra T[\rho]\xket$ is a Gaussian with standard
deviation $\l0$.
We will refer to $(1-\ph)$ as the {\it decoherence function} (what Gallis
and Fleming call the decorrelation factor). This function summarizes
all there is to know about the scattering environment, since
(\number\MasterEq) shows that it together with the Hamiltonian specifies the
dynamical behavior of our system entirely. 

Let us take a closer look at the last term.
In the position representation,
observation (IV) shows that it approaches $-\Lambda\rh$ far from the
diagonal. Near the diagonal, we can  use (V) to expand it as
$$\ybra- \Lambda(\rho-T[\rho])\xket \approx 
 -i\Lambda \vQ\cdot(\vy-\vx) 
  - \oh\Lambda(Q_m Q_n+S_{mn})(y_m-x_m)(y_n-x_n),$$ 
where we have dropped cubic and higher order terms in $|\vy-\vx|$, 
the distance to the diagonal.
Now since $\ybra[\vQ\cdot\vx,\rho]\xket = \vQ\cdot(\vy-\vx)\rho$, we can
absorb the first term into the Hamiltonian as a linear potential and write
(\number\MasterEq) as  
$$\dot\rho = -{i \over \hbar}[H+\Lambda\vQ\cdot\vx,\rho] - \Lambda
D[\rho],$$ 
$$\hbox{where }\quad <\vy|D[\rho]|\vx> \approx 
\oh (Q_m Q_n+S_{mn})(y_m-x_m)(y_n-x_n)\rh.$$ 

\noindent
This linear potential should come as no surprise --- it is simply the radiation
pressure term, and causes no dissipation.
(By dissipation we will mean a process that increases entropy, 
{\ie}, that converts pure states into mixed states. More quantitatively, we
will say that we have dissipation if the linear entropy $1-\tr\rho^2$
increases.)   From observation (III) we know that in general we do have
dissipation, so the cause of this must be the term $D[\rho]$.

To get a better feeling for what is happening, let us look at
what effect a single scattering would have in phase space. By transforming
Eq. (\number\RhoHitEq) to the Wigner representation 
[\Wigner=\refnr\rnr, \Kim=\refnr\rnr]
$$W(\vx,\vp) \equiv \ootpc \int 
 \rho(\vx+\vu/2,\vx-\vu/2)\,e^{i\vp\cdot\vu} d^3u$$
and doing some algebra, we see that a single scattering
has the effect
\WignerHitEq=\eqnr
$$W_f(\vx,\vp) = \int W_i(\vx,\vp-\vq)\,P(\vq)
d^3q,\eq$$ 
{\ie} a smearing out in momentum space.
Thinking of W as a probability distribution in phase
space, this convolution with the probability distribution
$P(\vq)$ corresponds to giving our particle a random
momentum kick. Thus we can use the central limit theorem
and approximate the effect of $n$ consecutive hits by an
equation identical to $(\number\WignerHitEq)$ but with $P(\vq)$ replaced by
a Gaussian with mean $n\vQ$ and covariance matrix $nS$. Since the exact
number of scatterings is not known but Poisson distributed 
with mean and variance equal to $n = \Lambda t$, the Gaussian will in fact
have the covariance matrix $\Lambda t (Q_m Q_n+S_{mn})$. Thus we see that
for the non-isotropic case $\vQ \not= 0$ the purely epistemological
uncertainty as to how many scatterings have occurred increases the
rate of wave function collapse, the rate of damping of off-diagonal elements. 

Returning to our density matrices, this
means that if $\Lambda$ is so large that there are many scatterings on
time scales shorter than that of ordinary Schr\"odinger evolution, then we
can replace the function $P$ in the master
equation $(\number\MasterEq)$ by a Gaussian with the same mean and
covariance matrix. Equivalently, we can replace $\ph$ by a Gaussian with the
same values of zeroth, first and second derivatives at the origin. This shows
that in the limit of large $\Lambda$, our Eq. $(\number\MasterEq)$ reproduces
the GRW equation  $(\number\GRWeq)$ {\it exactly}.
Furthermore, since the Gaussian is
the Green function of the diffusion equation, this means that we
can incorporate our dissipative term into the Wigner
equation {$[\number\Wigner,\number\Kim]$} as a simple diffusion term
of the type $\nabla^2_p$:
\FokkerPlanckEq=\eqnr
$$\eqalign{
\dot W = 
\biggl[
 &-{p_i \over m} \ddxi + 
\biggl({\partial V \over {\partial x_i}} - 
\Lambda Q_i\biggr)\ddpi\crr
&+ {1 \over \hbar}\sum_{n=1}^{\infty} {2^{-2n} \over (2n+1)!}
\biggl(\hbar\ddxi\ddpi\biggr)^{2n+1} V
+ D_{ij}\ddpi\ddpj
\biggr]\,W,
}\eq$$  
where $V(x)$ is the potential and the diffusion coefficient
$D_{ij}\equiv\Lambda(S_{ij}+Q_i Q_j)$. 
Here the spatial derivatives
in the infinite sum are to be understood to act only on $V$,
not on $W$. 
There are two well-known limits in which the Wigner
equation goes over into the Liouville equation of classical
statistical mechanics: when $V$ is at most quadratic, and when
$\hbar\to 0$. Because of our extra diffusion term, we get
yet a third classical limit:
in the limit of large $D_{ij}$, the
diffusive smoothing becomes so effective that it damps out
all the momentum-derivatives in the infinite sum, and
$(\number\FokkerPlanckEq)$ approaches the Liouville equation
with diffusion, an equation of Fokker-Planck type. 
This is yet another example of how macroscopic
objects start behaving classically, since as we will
soon see, $D_{ij}$ is roughly proportional to the size of our
object. Thus an object will evolve according to classical
dynamics if it has a strong interaction with its
environment. (The diffusion term in the resulting
Liouville equation is in no way a departure from classical
dynamics, since the Brownian motion due to random scatterings
must be taken into account also in a purely classically
analysis.)

\beginsection{4. SPECIFIC EXAMPLES}

In this section we will apply our results to specific scattering processes.
We will first calculate the shape and width of the decoherence function,
then study what the density matrix of a free particle converges to as
$t\to\infty$.

\bl\noindent
{\bf 4.1. The Decoherence Function}
\bl

Now let us evaluate $\vQ, S_{ij}$, $D_{ij}$ and the decoherence function
$1-\ph$ for some physically interesting cases. 
If the scattering cross section is given by
$\dsdo = \sigma\,\ftp$, where $\sigma$ is the total cross
section and the angular part $f$ is normalized so as to
integrate to unity, then $\pk(\vq) =
\delta(|\vq-\vk|-k)\,f/k^2$, where $k\equiv|\vk|.$
Using the properties of the Fourier
transform under translation and reflection and choosing
our coordinates so that $\vr = (0,0,r)$, we get
$$\eqalign{
\pkh(\vr) 
&= e^{-i\vk\cdot\vr}\int e^{iqr \ct}
\delta(q-k) \ftp \st d\theta d\varphi dq\crr
&= e^{-i\vk\cdot\vr}\int_0^1 e^{ikru} f(\arccos\,u,\varphi)
du d\varphi.
}$$
With a generic anisotropic radiation spectrum 
$\mu(\vk)$, the rate of
decoherence will be different along different spatial
directions and we will get radiation pressure. Since
neither of these two complications is particularly
illuminating, we will restrict ourselves to isotropic
radiation, {\ie} take ${\mu(\vk) = (4\pi\vk^2)^{-1}
\l0\nu(\l0|\vk|)}$, where the spectrum $\nu$ is a probability
distribution on the positive real line and $\l0$ is some 
typical wavelength. Performing the angular integration, this yields
\PhatrEq=\eqnr
$$\ph(\vr) = \int_0^{\infty}g(ur/\l0)\nu(u)du,\eq$$
where
$$g(x) \equiv \sxx \int_0^1 e^{ixu}
f(\arccos\,u,\varphi) du d\varphi.$$
For the case of photon scattering against both a free charge and a
dielectric sphere much smaller than the photon
wavelength, we get \Jackson=\refnr
{\ref} the angular dependence 
\gOneEq=\eqnr
$$\ftp = {3 \over {16\pi}}(1+\cos^2\theta),$$
which
yields
$$g(x) = g_1(x)\equiv
{3 \over 2} \biggl[\cx+(x^2-1)\sxx\biggr]
{\sin^2x \over {x^4}}.\eq$$
We will refer to this mixture of S-waves and D-waves as SD-wave scattering.
Another physically important case is pure S-wave scattering, {\ie}
\gTwoEq=\eqnr
$$\ftp = {1 \over {4\pi}},\quad\hbox{which
yields}\quad g(x) = g_2(x)\equiv
\biggl(\sxx\biggr)^2.\eq$$
This case applies among other things to an opaque spherical object of radius
$a$ much larger than ${\leff}$, which we can use to model say a dust particle
scattering optical photons. Here
the total cross section is frequency independent, roughly equal to the
geometrical cross section $\pi a^2$, and perhaps surprisingly, the
scattering amplitude turns out to be the same in all directions.

Since we are restricting ourselves to isotropic
radiation, we simply have 
the mean {$\vQ=0$} and the covariance matrix is
proportional to the identity matrix, {\ie} $S_{ij} =
s^2\delta_{ij}$, so all we need to calculate is the
standard deviation $s$. From Eq. (\number\PhatrEq) we get
\sEq=\eqnr
$$\ph''(0) = \l0^{-2}\izi\nu(x)x^2g''(0)dx = 
g''(0)\ex2,$$
so 
$$s = \l0^{-1}[-g''(0)]^{1/2}\ex2^{1/2},\eq$$ 
since the variance of a
probability distribution of zero mean is the negative of the second
derivative of its Fourier transform at the origin. For the
functions in (\number\gOneEq) and (\number\gTwoEq), the values we need are
$-g''_1(0) = 11/15$ and  $-g''_2(0) = 2/3$. Thus apart from these
numerical parameters depending only on the angular
part of scattering cross section, we see that the standard
deviation of the momentum kick is simply a certain
spectrally averaged momentum. 

Let us
define the  {\it effective wavelength} as 
$$\leff \equiv 1/s = \l0[-g''(0)]^{-1/2}\ex2^{-1/2}.$$ 
Then the diffusion matrix
$D_{ij} = \Delta\delta_{ij}$, where the scalar diffusion
coefficient  $\Delta\equiv\Lambda/\leff^2.$
Sunlight on earth, the $300K$ radiation from our surrounding, the
cosmic microwave background radiation and the cosmic neutrino background
all have Planck spectra, corresponding to temperatures of roughly
$5800K$, $300K$, $2.7K$ and $2.0K$, respectively.
For a Planck spectrum, we have 
$$\nu(x) = \toz3 {x^2 \over e^x - 1},$$
where $\zeta(s) \equiv
\soi n^{-s}$ is the Riemann Zeta function.
For this particular case, $\ex2 = 4!\zeta(5)/2!\zeta(3)$, 
so we get $\leff\approx
0.381\l0$ for the S-wave case and $\leff\approx
0.363\l0$ for the SD-wave case. For a point spectrum $\nu(x)=\delta(x-1)$
with a single wavelength $\l0$ we simply get $\ex2 = 1$, so the
corresponding values are $\leff\approx 1.225\l0$ 
and $\leff\approx 1.168\l0$.

Now let us calculate the decoherence function $1-\ph$ for the Planck case. 
By expanding $\nu$ as a geometric series, we get
$$\ph(\vr) = \izi g(ur/\l0)\nu(u)du = 
{2 \over \z3}\soi \izi g(xu)u^2e^{-nu}du.$$
For the S-wave case $g=g_2$, this integral can be
done elementarily, yielding
$$\ph(\vr) = {1 \over \z3} 
\soi n^{-3}\left[{1+{\left({2r/\l0 \over n}
\right)}^2}\right]^{-1}.$$
Numerical integration is required in the
case $g=g_1$, and gives a decoherence function $1-\ph(r/\leff)$ that differs
by less than 1\% from the S-wave case when appropriately rescaled. This
decoherence function is plotted for Figure 1, together with 
the S-wave and SD-wave decoherence functions for point spectra. Also plotted
is the GRW decoherence function, for which $\ph$ is Gaussian.

For dielectric spheres with $a\ll \leff$ and frequencies well below any
resonances, the cross section depends on frequency according to Rayleigh's
$k^4$-law, so we must replace the Planck spectrum by $\nu(x) = {6! \over
\zeta(7)} {x^6 \over e^x - 1}$. 

\eject
{\noindent\bf}
\vglue8.2truecm
\noindent
{\bf Figure 1.} Decoherence functions $(1-\ph)$ for different spectra $\nu$
and angular distributions $f(\theta,\varphi)$. The functions have been scaled so as to all have second derivative unity at the origin. 

\bl

Let us define the {\it coherence time} $\tau\equiv\Lambda^{-1}$.
GRW have given an exact solution of Eq. $(\number\MasterEq)$ in
{$[\number\GRW]$} for  the case of
a free particle, so to get a feeling for what happens we will only
mention the simple case when $\tau$ is is much shorter than the dynamical
timescale, so that we can neglect the ordinary Schr\"odinger evolution
for
short times. Thus setting $H \approx 0$ and using the Gaussian
approximation that is valid for $t\gg\tau$, Eq. $(\number\MasterEq)$ has the
short-time solution $$\rho(\vx,\vy,t_0+t) \approx 
 \rho(\vx,\vy,t_0) e^{-\Lambda t\left(1-e^{-r^2/2\leff^2}\right)}.$$ 
Thus we see that
far from the diagonal, for $|\vy-\vx| \gg \leff$, the elements of the
density matrix are damped out as $e^{-\Lambda t}$ independently of $\leff$.
Near the diagonal, on the other hand, for $|\vy-\vx| \ll \leff$,
the crucial parameter is the diffusion parameter $\Delta =
\Lambda/\leff^2$, since the damping goes as 
$e^{-\Delta |\vy-\vx|^2 t/2}$.

Table $1$ gives $\leff, \phi$ and $\tau$ for a variety of
radiation sources and Table $2$ gives the diffusion parameter $\Delta$ for
the center-of-mass of three different objects, in order of decreasing
strength. 

\bl
\goodbreak\nobreak

{\vbox{
\raggedright
\noindent
{\bf Table 1.} Properties of various scattering processes
\tabskip = 1em
\halign{#&$#$&$10^{#}$&$#$\cr
\streck
Cause of collapse&
\omit $\leff$&
\omit $\phi[\cm^{-2}\s^{-1}]$&
\omit $\tau_{electron}$\cr 
\streck
300K air at 1 atm pressure&0.1\,\A&24&10^{-13}\s\cr
300K air in lab vacuum&0.1\,\A&11&1\,\s\cr
Sunlight on earth&900\,\nm&17&6\,{\rm months}\cr
300K photons&0.02\,\mm&19&1\,{\rm day}\cr
Background radioactivity&10^{-14}\m&-4&10^{11} \yrs\cr
Quantum gravity&1\,\km-10^{10}\m&109&30\s\cr
GRW effect&100\,\nm&\omit n/a&10^9\yrs\cr
Cosmic microwave background&2\,\mm&13&10^4\yrs\cr
Solar neutrinos&0.1\,\A&11&10^{26}\,\yrs\cr
Cosmic background neutrinos&3\,\mm&13&10^{44}\yrs\cr
\streck
}}}

\bl
\goodbreak
\nobreak
%\hglue0.3truecm{\vbox{
{\vbox{
\raggedright
\noindent
{\bf Table 2.} 
Decoherence rate $\Delta$ in $\cm^{-2}\s^{-1}$ for various objects

\noindent
and scattering processes 

\tabskip = 1em
\halign{#&$10^{#}$&$10^{#}$&$10^{#}$\cr
\streck
Cause of apparent&
\omit Free&
\omit $10\mu m$&
\omit Bowling\cr 
wave function collapse&
\omit electron&
\omit dust&
\omit ball\cr 
\streck
300K air at 1 atm pressure&31&37&45\cr
300K air in lab vacuum&18&23&31\cr
Sunlight on earth&1&20&28\cr
300K photons&0&19&27\cr
Background radioactivity&-4&15&23\cr
Quantum gravity&-25&10&22\cr
GRW effect&-7&9&21\cr
Cosmic microwave background&-10&6&17\cr
Solar neutrinos&-15&1&13\cr
\streck
}}}

\bl
\goodbreak

The decoherence rates for photons and air agree well with those given by
Joos and Zeh $[\number\JoosZeh]$.
The effect of air molecules is seen to dominate at room
temperature not only at atmospheric pressure but also in a laboratory vacuum
of $10^6$ particles/cm$^3$. The radioactivity figures are quite crude,
since the fluxes of $\alpha$, $\beta$ and $\gamma$ rays vary widely with
location and  surrounding \ref. The energy
of the free electron matters only in the case of air [\number\Jackson], where
it has been taken to be 1 keV.
The reason that the cosmic background neutrinos {\ref} are
completely negligible compared to those from the sun {\ref} despite similar
fluxes is that the weak scattering cross section falls off as energy squared
in our regime of interest, which is well below the W mass of 81 GeV. 
The neutrino effect is completely impossible to shield against - a
typical neutrino coasts undisturbed straight through our planet. 

Two sets of numbers from
other sources have been put in for comparison. 
The GRW
values were chosen ad hoc in {[\number\GRW]} to match observation as well as
possible. The quantum gravity values from 
[\Ellis=\refnr\rnr,\Hawking=\refnr\rnr] are based on
dimensional analysis and several perhaps questionable assumptions. For
instance, $\leff$ is assumed to depend on the mass of the particle, ranging
from about 1 km for a proton up to astronomical $10^{10}$ m for an
electron.

The main observation to make about the two latter effects is
that although they have been highly publicized, they rank only fifth and
sixth in strength, trailing by more than twenty orders of magnitude.
Hence an experiment devised to measure them would have to be nearly
perfectly shielded from all the stronger sources of decoherence.
Blocking out optical photons should pose no problem.
Background radioactivity could conceivably be controlled by using
ultra-pure equipment and performing experiments in a deep mine, thereby
avoiding virtually all cosmic rays except muons. The effect of air (or any
other surrounding substance) and blackbody radiation from the surrounding is
strongly temperature dependent (typically $\Delta \propto T^5$), and can
hence be reduced by nine orders of magnitude by working at liquid Helium
temperatures. Even under such conditions, they would still be stronger
than the GRW and quantum gravity effects.
Although it may become feasible to cool a macroscopically large apparatus
to microkelvin temperatures, environmentally induced decoherence is in a
sence endemic: In order to observe the object of an experiment, we
must by definition let it interact with something else.

Of no small importance is that we know
that all effects in the table except those of quantum gravity and GRW
do in fact occur.  Because of this,
a number of experiments that have been proposed are likely to yield
inconclusive results.
For example, Hawking $[\number\Hawking]$ and others have conjectured that
quantum gravity effects might be able to explain the apparent collapse of the
wave function. Ellis-Mohanty-Nanopoulos have made the estimates of such
wormhole effects quoted above $[\number\Ellis]$, but apart from the fact that
there is no experimentally tested theory of quantum gravity, these effects
would probably be impossible to detect if an attempt where made to measure
them, since they would so to speak drown in environmental noise.
In fact, calculations in {$[\number\JoosZeh]$} show that even ordinary
Newtonian gravity often has a stronger decoherence effect than quantum
gravity.

The same goes for suggestions to measure an independent GRW effect. 
Squires [\number\Squires] suggests that a GRW effect might indeed exist and
be  caused by some yet unknown physics that he speculates might be ``the
physics of the 21st century". Our results have shown that the physics of this
century produces an almost identical effect, and that an additional ``new
physics" contribution of the magnitude postulated by GRW would probably 
be too many orders of magnitude weaker than the decoherence effect to be
detectable.  

Also Rae [\number\Rae] suggests that the GRW effect might be caused by new
physics, and speculates that 
${\Lambda}$ and ${\alpha}$ might be new constants of nature. Our
results have shown that when decoherence dominates, the ${\Lambda}$ and
${\lambda}$ that would be measured in a GRW experiment would be calculable
from scattering cross sections and spectra of scattering particles.

The fact that decoherence and the GRW effect have an almost indistinguishable
impact on the density matrix opens up an interesting possibility: if a 
GRW effect due to new physics indeed exists, then it might be much stronger
than originally postulated without contradicting our experience. Thus
experiments devised to search for a GRW effect are by no means without
interest.

\bl\noindent
{\bf 4.2. Coherence Lengths}
\bl

What is the width $\Dx$ of the wavepacket for a free
particle? Any textbook will give the answer that 
$\Dx\to\infty$ as $t\to\infty$, but we are now in a
position to give a more subtle and indeed finite
answer.

In the presence
of our scattering, $\Dp\to\infty$ like $\hbar\sqrt{Dt}$
just as for Brownian motion in momentum space. 
For a truly free particle, $\Dp$ remains constant
whereas $\Dx\to\infty$ like $t\Dv=t\Dp/m$. 
An often overlooked fact is that there is nothing
``quantum'' about this whatsoever, other than that the
uncertainty principle prohibits $\Dp = 0$ initially. Since
the free particle Hamiltonian is
quadratic, the Wigner equation (\number\FokkerPlanckEq)
reduces to the Liouville equation and the increase in
$\Dx$ only reflects a classical type of uncertainty, our
ignorance. 

Let us keep the conventional
definition
$$\Dx\equiv
\left[{\left<x^2\right>-\left<x\right>^2}\right]^{1/2}$$
for pure states, but redefine $\Dx$ for mixed states to be 
the {\it coherence
length}, roughly speaking the largest distance from the diagonal where the
spatial density matrix has non-negligible components.
Let us redefine $\Dp$ analogously. 

More formally, let us consider a density matrix $\rho$ that
is an incoherent mixture of tiltedly squeezed states (Gaussians in Wigner
phase space), all of which have the same values of the $\Dx$ and $\Dp$ but
with different $\left<x\right>$, $\left<p\right>$ and 
$\left<xp\right>$. GRW
analyze the solution to Eq. $(\number\GRWeq)$  in detail in [\number\GRW], 
and show that such a $\rho$ can be expanded as such a mixture for all times,
but with $\Dx$, $\Dp$ and the mixing function changing with time.  During the
undisturbed Schr\"odinger time evolution between
scatterings, $\Dx$ increases as usual while $\Dp$ remains
constant (a shearing in phase space).  A scattering causes $\Dx$ to decrease
abruptly while $\Dp$ can either increase or decrease.
The net result of the interplay between these two
counteracting effects is that although the conventional
uncertainties become infinite, both $\Dx$ and $\Dp$
converge to finite limiting values as $t\to\infty$. We
might interpret this as that after a long time
our particle definitely is in a state with spreads $\Dx$ and
$\Dp$, but we have absolutely no knowledge as to where in
phase space this state is centered.

In our notation, GRW show that the limiting
values of the spreads are 
$$\Dx = u\,\leff\quad\hbox{and}
\quad\Dp = v\,{m\leff\over\tau},$$ 
where
$$v\equiv \eta^4{1+u\over2u},\quad 
\eta \equiv \left({\hbar\tau\over m\leff^2}\right)^{1/2},$$
and $u$ is the positive solution to
$$4u^3 = 2\eta^2u\sqrt{1+2u}+\eta^4(1+u)(1+2u).$$
The dimensionless constant $\eta$ ranges from $10^{-29}$ to
$10^{26}$ for the examples in the tables above.
GRW only consider the macroscopic limit $\eta\ll 1$, where
$$\Dx\approx 
2^{-1/4} \left({\hbar\over m\Delta}\right)^{1/4}
\quad\hbox{and}\quad \Dp\approx 
2^{-1/4} \left({\hbar^3 m \Delta}\right)^{1/4},$$
but we will also be interested in the microscopic limit 
$\eta\gg 1$, where we get 
$$\Dx\approx 
2^{-1/2}{\hbar\tau\over m\leff}
\quad\hbox{and}\quad \Dp\approx 
2^{-1/2} {\hbar\over\leff}.$$
Table 3 contains the limiting value of $\Dx$ for
the previously discussed objects and decoherence
sources. (These figures are to be interpreted as upper
limits to the true coherence length, since in reality the
different effects all contribute separately.)
``n/a" has been entered
for the cases where $\tau$ is greater than the age of the
universe, so that  the system has not yet had time to
approach the limiting value of $\Dx$.

It is interesting to compare the uncertainty products
$\Dx\Dp$ from above with the minimum value $\hbar/2$
allowed by the uncertainty principle. 
For the macroscopic case $\eta\ll 1$, we get
$${\Dx\Dp\over \hbar/2} \approx \sqrt 2,$$
whereas the 
microscopic case $\eta\gg 1$ yields
$${\Dx\Dp\over \hbar/2} \approx \eta^2.$$
This indicates that we should expect to find macroscopic
systems such as dust particles in states that are nearly
minimum uncertainty states, but microscopic systems
in highly tiltedly squeezed states where the uncertainty
product is much larger than its minimum. For a free electron
decohered only by 300K photons, for instance, 
$\Dx\Dp \approx 10^{11}\hbar$.
This conclusion is likely to be valid quite generally, 
also when the initial state is not of the form assumed above,
since recent work by Zurek, Habib and Paz {\ref} has indicated that quite
general states tend to approach generalized coherent states (states with a
Gaussian Wigner function) when they interact with their environment.

\goodbreak
\bl
\nobreak
{
\raggedright
\noindent
{\bf Table 3.} 
Coherence lengths $\Dx$ 
caused by various decoherence sources
\tabskip = 1em
\halign{#&$10^{#}\,\m$&$10^{#}\,\m$&$10^{#}\,\m$\cr
\streck
Cause of apparent&
\omit Free&
\omit $10\mu m$&
\omit Bowling\cr 
wave function collapse&
\omit electron&
\omit dust&
\omit ball\cr 
\streck
300K air at 1 atm pressure&-6&-17&-21\cr
300K air in lab vacuum&7&-13&-18\cr
Sunlight on earth&9&-12&-17\cr
300K photons&4&-12&-16\cr
Background radioactivity&\omit n/a&-11&-15\cr
Quantum gravity&4&-9&-15\cr
GRW effect&19&-9&-15\cr
Cosmic microwave background&10&-8&-14\cr
Solar neutrinos&\omit n/a&\omit n/a&-13\cr
\streck
}}
\bl
\goodbreak

This paper has focused entirely on scattering. The case of coupled harmonic
oscillators has been studied by
numerous authors, and {$[\number\JoosZeh]$} has analyzed interactions of
Coulomb type. However, many other interesting sources of decoherence still
remain to be analyzed in detail, and might well turn out to be stronger than
any of those discussed above.

%\beginsection{5. DECOHERENCE AND THE INTERPRETATION OF QUANTUM MECHANICS}
\bl\bl\goodbreak\noindent
{\bf 5. DECOHERENCE AND THE INTERPRETATION OF}

\noindent
{\bf QUANTUM MECHANICS}
\bl

In a discussion of density matrices \ref, Feynman writes:
``When we solve a
quantum-mechanical problem, what we really do is
divide the universe into two parts - the system in which we are
interested and the rest of the universe. We then usually act as if
the system in which we are interested comprised the entire
universe."
In this spirit we summarize our scattering results:
The effects of ``the rest of the universe" can be incorporated into two
additional terms in the Heisenberg equation of motion; a radiation-pressure
term that can be absorbed into the Hamiltonian and a dissipative term that
transforms pure states into mixed states. For macroscopic systems, the former
is usually negligible whereas the latter effectively damps out spatial
superpositions.

As mentioned in the introduction, the decoherence effect removes
a serious deficiency {\ref} from non-collapse theories. This makes possible a
self-consistent interpretation of quantum mechanics that might be called
the {\it many decohering worlds interpretation}:

\bl

\qlist{
{\bf Q}: Why do we {\it experience} collapse?
}

\qlist{
{\bf A}: Because as shown in $[\number\Everett]$, collapse and QM
probabilities will be experienced by almost all observers in the grand
superposition.
}

\qlist{
{\bf Q}: Why do we only experience macroscopically ``nice"
superpositions?
}

\qlist{
{\bf A}: Because all others get damped out by decoherence effects before we
have time to observe them.
}

\qlist{
{\bf Q}: Does the Von Neumann reduction have anything to do with mental
processes and our observing the system? 
}

\qlist{
{\bf A}: No, it has already occurred by the time we observe the object.
}

\qlist{
{\bf Q}: What about Schr\"odinger's cat?
}

\qlist{
{\bf A}: Due to decoherence, it is for all practical purposes either dead or
alive - we just don't know which. 
}
(Several of the quoted decoherence authors would disagree with these
answers.) 

\bl

If there is agreement on any philosophical interpretation of our century's
developments in physics, it is that the ultimate reality is more bizarre than
anything we ever dreamed of. Hence we no longer feel that we can reject a
theory merely on the grounds that it involves bizarre and counter-intuitive
notions. Nonetheless, most people would probably agree that in order for any
theory to be hailed as fundamental, it should satisfy the following minimum
requirement: {\it it should be able to explain why we do not experience any of
the bizarre notions that it introduces.}

In the case of quantum mechanics, this would entail using first
principles to explain why the wave function {\it seems} to collapse and why
the macroscopical world {\it seems} to obey the laws of classical physics.
Despite Bohr's many examples of classical correspondence for high-energy
eigenstates, despite Schr\"odinger's invention of coherent states and despite
the well-known classical limit when $\hbar \to 0$, this minimum requirement
was not met until the work on decoherence in the last two
decades. Until then, nobody had explained why we never experience say
spatial superpositions of macroscopic objects (without invoking new physics
or superselection rules postulated ad hoc).

While recognizing this success, it is important to remember
that although quantum mechanics including the decoherence effect meets our
minimum requirement, this is also all it does.  
Although the existence of environmentally induced decoherence does
explain why we never experience bizarre macrosuperpositions, and thus makes
dynamic reduction mechanisms (DRMs) such as those proposed by GRW
and {$[18,35,36]$\nextref\nextref}
unnecessary for that purpose, it does no more than
that. In other words, despite decoherence, macrosuperpositions still do exist -
decoherence merely explains why we cannot experience them. Instead of being
destroyed, superpositions spread to ever larger subsystems of the universe as
everything gets more and more quantum-mechanically entangled with everything
else.

The unease with which many authors view this
state of affairs was one of the main motivations 
in introducing DRMs. 
In DRMs, this perpetual entanglement is avoided by making the additional 
postulate that the wave function itself really does get localized, usually
in discrete, collapse-like jumps.
The price is that DRMs need to postulate new physics
without explaining any phenomena that decoherence alone does not, thus leaving
themselves vulnerable to Occam's razor.

As Dieks {\ref} and others have pointed out, there is also a second source of
unease.
This is that a density matrix describing a mixed state can be
expressed as a statistical mixture of pure states in 
infinitely many different ways. 
Thus we
are not justified to make the interpretation that an
object ``really" is at a definite position, even if decoherence
has made its spatial density matrix diagonal. 

But then again, tell one of your friends that the world is a
weird place, and the answer will be: ``So what else is new?"

\beginsection{ACKNOWLEDGEMENTS}

The author would like to thank J. Bendich, E. Bunn,
A. Elby, and L. Yeh for
illuminating discussions on the subject of this paper, and Prof. G. Ghirardi
and Prof. H. Shapiro for many useful comments.

\frenchspacing
\beginsection{REFERENCES}
%\narrower

\refnr = 1
\smallskip
\refn G. C. Ghirardi, A. Rimini and T. Weber, {\it Phys. Rev. D}
{\bf 34}, 470 (1986).

\refn H. Everett III, {\it Rev. Mod. Phys.} {\bf29}, 454 (1957).

\refn H. Everett III, {\it The Many-Worlds Interpretation of
Quantum Mechanics}, B. S. DeWitt and N. Graham, eds. 
(Princeton University Press, Princeton, 1986).

\refn J. A. Wheeler, {\it Rev. Mod. Phys.} {\bf 29}, 463 (1957).

\refn L. M. Cooper and D. van Vechten, {\it Am. J. Phys.} {\bf
37}, 1212 (1969).

\refn L. N. Cooper, ``Wave function and observer in quantum
theory", in {\it The Physicist's Conception of Nature}, J. Mehra,
ed. (Reidel, Dordrecht, 1983).

\refn B. S. DeWitt, {\it Phys. Today} {\bf 23} (9), 30 (1971).

\refn H. D. Zeh, {\it Found. Phys.} {\bf 1}, 69 (1970).

\refn O. K\"ubler and H. D. Zeh, {\it Ann. Phys.} {\bf 76}, 405
(1973).

\refn E. Joos, {\it Phys. Rev. D} {\bf 29}, 1626 (1984).

\refn E. Joos and H. D. Zeh, {\it Z. Phys. B} 
{\bf 59}, 223 (1985).

\refn S. Machida and N. Namiki, {\it Prog. Theor. Phys.} {\bf 63},
1457, 1833 (1980).

\refn W. H. Zurek, {\it Phys. Rev. D}
{\bf 24}, 1516 (1981); {\bf 26}, 1862 (1982).

\refn W. G. Unruh and W. H. Zurek, {\it Phys. Rev. D}
{\bf 40}, 1071 (1989).

\refn W. H. Zurek, {\it Phys. Today} {\bf 44} (10), 36 (1991).

\refn M. Cini, {\it Nuovo Cimento B} {\bf 73}, 27 (1983).

\refn E. Peres, {\it Am. J. Phys.} {\bf 54}, 688 (1986).

\refn P. Pearle, {\it Phys. Rev. A} {\bf 39}, 2277 (1989).

\refn M. H. Partovi, {\it Phys. Lett. A} {\bf 137}, 445 (1989).

\refn M. R. Gallis and G. N. Fleming, {\it Phys. Rev. A} {\bf 42}, 38 (1989).

\refn E. J. Squires, {\it Phys. Lett. A} {\bf 158}, 431 (1991).

\refn A. I. M. Rae, {\it J. Phys. A} {\bf 23}, L57 (1990).

\refn K. Kraus, {\it States, Effects and Operations} (Springer,
Berlin, 1983).

\refn E. P. Wigner, {\it Phys. Rev.} {\bf 40}, 749 (1932).

\refn Y. S. Kim and M. E. Noz, {\it Phase Space Picture of Quantum
Mechanics: Group Theoretical Approach} (World Scientific,
Singapore, 1991).

\refn J. D. Jackson, {\it Classical Electrodynamics}
(Wiley, New York, 1975), p.414.

\refn {\it Appl. Ratiat. Isot.} {\bf 39}, 717 (1988).

\refn E. Kolb and M. S. Turner, {\it The Early Universe}
(Addison-Wesley, Reading, 1990).

\refn J. N. Bahcall and R. K. Ulrich, 
{\it Rev. Mod. Phys.} {\bf 60}, 297 (1988);
{\bf 59}, 505 (1987).

\refn J. Ellis, S. Mohanty and D. V. Nanopoulos, 
{\it Phys. Lett. B}
{\bf 221}, 113 (1989).

\refn S. W. Hawking, {\it Commun. Math. Phys.} {\bf 87}, 395 (1982).

\refn W. H. Zurek, S. Habib and J. P. Paz, {\it Phys. Rev. Lett.} {\bf 70},
1187 (1993).

\refn  R. P. Feynman, {\it Statistical Mechanics}
  (Benjamin, Reading, 1972).

\refn Y. Ben-Dov, {\it Found. Phys.} {\bf 3}, 383 (1990).

\refn G. C. Ghirardi, P. Pearle and A. Rimini, 
{\it Phys. Rev. A} {\bf 42}, 78 (1990).

\refn G. C. Ghirardi, R. Grassi and A. Rimini, 
{\it Phys. Rev. A} {\bf 42}, 1057 (1990).

\refn D. Dieks, {\it Phys. Lett. A} {\bf 142}, 439 (1989).

\end